\documentclass{sig-alternate}
\usepackage{graphicx}
\usepackage{url}
\usepackage{alltt}
\usepackage{times}
\usepackage{mdwlist} 

\def\sharedaffiliation{
\end{tabular}
\begin{tabular}{c}}

\begin{document}

\conferenceinfo{WSDM'10,} {February 4--6, 2010, New York City, New York, USA.} 
\CopyrightYear{2010}
\crdata{978-1-60558-889-6/10/02} 

\title{Folks in Folksonomies:\\
Social Link Prediction from Shared Metadata}

\numberofauthors{5}
\author{
\alignauthor Rossano Schifanella$^{1}$\titlenote{\small Corresponding author. Email: \texttt{schifane@di.unito.it}. A good portion of the work in this paper was carried out 
while Dr. Schifanella was a visiting scholar at the Center for Complex Networks and 
Systems Research (CNetS) of the Indiana University School of Informatics and Computing.}
\alignauthor Alain Barrat$^{2,3}$
\alignauthor Ciro Cattuto$^{3}$
\and
\alignauthor Benjamin Markines$^{4}$
\alignauthor Filippo Menczer$^{3,4}$
\sharedaffiliation
\affaddr{$^1$ Department of Computer Science, University of Turin, Italy}\\
\affaddr{$^2$ Centre de Physique Th\'eorique (CNRS UMR 6207), Marseille, France}\\
\affaddr{$^3$ Complex Networks and Systems Laboratory, ISI Foundation, Turin, Italy}\\
\affaddr{$^4$ School of Informatics and Computing, Indiana University, Bloomington, IN, USA}
}

\maketitle

\begin{abstract}
Web 2.0 applications have attracted a considerable amount of attention
because their open-ended nature allows users to create light-weight semantic scaffolding
to organize and share content. To date, the interplay of the social and semantic
components of social media has been only partially explored. Here we focus on Flickr 
and Last.fm, two social media systems in which we can relate the tagging activity 
of the users with an explicit representation of their social network.
We show that 
a substantial level of local lexical and topical alignment is observable among users
who lie close to each other in the social network. We introduce a null model
that preserves user activity while removing local correlations, allowing 
us to disentangle the actual local alignment between users from statistical effects
due to the assortative mixing of user activity and centrality in the social network. 
This analysis suggests that users with similar topical interests are more likely to be 
friends, and therefore semantic similarity measures among users based solely on their 
annotation metadata should be predictive of social links. 
We test this hypothesis on the Last.fm data set, confirming that the social network 
constructed from semantic similarity captures actual friendship more accurately than  
Last.fm's suggestions based on listening patterns. 
\end{abstract}

{\small

}



%

\section{Introduction}
\label{sec:intro}

Social networking systems like Facebook and systems for content organization 
and sharing such as Flickr and Delicious have created information-rich ecosystems
where the cognitive, behavioral and social aspects of a user community
are entangled with the underlying technological platform.
This opens up new ways to monitor and investigate a variety of processes
involving the interaction of users with one another, as well as
the interaction of users with the information they process. 
Social media supporting tagging~\cite{mathes04folksonomies,hhls05social} are especially interesting
in this respect because they stimulate users to provide light-weight semantic
annotations in the form of freely chosen terms. Usage patterns 
of tags can be employed to monitor interest, track user attention, 
and investigate the co-evolution of social and semantic networks. 

While the emergence of conventions and shared conceptualizations
has attracted considerable 
interest~\cite{staab2002_emergent,mika07ontologies,steels2008,golder2006structure}, 
little attention has been devoted so far to relating, 
at the microscopic level, the usage of shared tags with the social links 
existing between users. 
The present paper aims at filling this gap. To this end we focus on Flickr and Last.fm,
as to our knowledge they are currently the only popular social media system where:
(1) a significant fraction of the users provide tag metadata for their content (photographs 
or songs), and (2) an explicit representation of the social links between users is readily 
available. 

The main question that we address in this study is the following:
given two randomly chosen users, how does the alignment
of their tag vocabularies relate to their proximity on the social network?
That is, does lexical alignment exist between neighboring users,
and if so, how does this alignment fade when considering users
lying at an increasing distance on the social graph? 
And if indeed such a relationship exists, does it allow us 
to predict social links from analysis of the semantic similarity 
among users, extracted from their annotations? 

\subsection*{Contributions and outline}

The main contributions of this paper are summarized as follows:

\begin{itemize}

\item In \S~\ref{sec:hetero} we show that strong correlations exist across several 
measures of user activity, and characterize the mixing patterns that involve 
user activity and user centrality in the social network. 

\item In \S~\ref{sec:alignment} we develop sound measures of tag overlap.
We further introduce appropriate null models to disentangle the actual
local alignment between users from statistical effects due to the
mixing properties of user activity and centrality in the social
network. We apply these measures to the Flickr and Last.fm data
sets. The resulting analysis shows that, despite neither Flickr nor
Last.fm support globally-shared tag vocabularies, a substantial level
of local lexical (shared tags) and topical (shared groups) alignment
is observable among users who are close to each other in the social
network.  We also find that some observables are more adequate than
others to measure lexical and topical alignment, in the sense that
they are less sensitive to purely statistical effects.

\item In \S~\ref{sec:predict} we inquire if the observed correlations between 
annotation metadata and social proximity allow to use semantic similarity between 
user annotations as statistical predictors of friendship links. We evaluate a 
number of semantic similarity measures from the literature, based on 
Last.fm metadata.  We find that when we consider the annotations of the most active 
users, almost all of the semantic similarity measures considered 
outperform the neighbor suggestions from the Last.fm system at predicting actual 
friendship relations. Scalable semantic similarity measures such as 
Maximum Information Path, proposed by some of the authors, are among those  
achieving the best predictive performance. 

\end{itemize}


\section{Related work}
\label{sec:related}

One of the first quantitative studies on Flickr is presented by Marlow \textit{et al.}~\cite{ht06marlow}, who discuss the heterogeneity of tagging patterns and perform
a preliminary analysis of vocabulary overlap between pairs of users. The analysis 
suggests that users who are linked in the Flickr social network have on average
a higher vocabulary overlap, but no assessment is made of biases and other
correlations that could be responsible for the reported observation.

The structure and the temporal evolution of the Flickr social network are investigated
in several papers~\cite{kumar06kdd,mislove-2007-socialnetworks,mislove-2008-flickr}. 
Leskovec \textit{et al.}~\cite{kumar08kdd} place a special emphasis on the local 
mechanisms driving the microscopic evolution of the network.

The role of social contacts in shaping browsing patterns on Flickr has also 
been explored~\cite{lerman06flickr,vanzwol07flickr}, providing insights into the behavior
and activity patterns of Flickr users.

Prieur \textit{et al.}~\cite{prieur08cooperation} investigate the role of Flickr groups
as coordination tools, and explore the relation between the density of the social
network and the density of the network of tag co-usage among the group members.

Liben-Nowell and Kleinberg~\cite{libennowell2003cikm} explore several notions 
of node similarity for link prediction in social networks. In our own prior work 
we performed a systematic analysis of a broad range of semantic similarity measures 
based on folksonomies, that can be applied directly to build networks of 
users, tags, or resources~\cite{Markines08HT,Markines09www,ht09_GaL_MIP_poster}. 
Here we build upon this evaluation framework.

Li \textit{et al.}~\cite{LiWWW08} propose a system to discover common 
user interests and cluster users and their saved URLs by different interest topics.
They use a Delicious data set to define implicit links between users based 
on the similarity of their tags. However they do not correlate the interest 
clusters with social connections.

Perhaps the work that most directly relates to our approach is by Santos-Neto 
\textit{et al.}~\cite{Santos-NetoHT09}, who explore the question of whether 
tag-based or resource-based interest sharing in tagging systems relate to 
other indicators of social behavior. 
The authors analyze the CiteULike and Connotea systems, which 
deal with scientific publication and lack explicit social network components.  
Therefore they are unable to directly explore social friendship between two 
users, and instead look at participation in the same discussion group, with 
mixed results. They do not find a statistical correlation 
between the intensity of interest sharing and the collaboration levels. 
Our present results are both more explicit (we deal with pairs of users 
rather than groups) and more conclusive. Furthermore 
we are able to evaluate our interest-based predictions against external 
suggestions based on independent data, quantifying the applicative value of our findings. 


\section{Data sets}
\label{sec:data}

Flickr makes available most of its public data
by means of API methods (\url{flickr.com/api}).
The data used for the present study were obtained by using the public Flickr API
to perform a distributed crawl of the content uploaded to Flickr
between January 2004 and January 2006. The system was crawled during
the first half of 2007. The crawling task was distributed by dividing
the above interval of time into work units consisting of smaller time windows,
and crawling each time window separately. Each crawler was programmed
to issue search queries for every known tag, limited to its temporal window of competence,
as well as to issue search queries for photos uploaded in the same interval.
As new tags were discovered, they were added to a global table shared by all the
crawlers.
Separate crawls were performed to explore the Flickr social network 
(in Flickr jargon, the ``contacts'' of a given user are her nearest neighbors
in the social network, as represented in the system),
as well as group membership information.

Overall, the data set we analyzed comprises $241,031$ users for whom we have
tagging information, and $118,144$ users for whom we also have 
group membership information.
%
Our analysis will focus on two networks. The first one, $G_0$, comprises
the Flickr users for whom we have tag, group and contact information.
It consists of $118,144$ nodes (users) and $2,263,182$ edges (contacts between users).
The second network, $G_1$, is obtained by extending $G_0$
to include all the neighbors of its nodes, neighbors for whom we may not have
tagging, group membership, or complete contact information.
$G_1$ comprises $983,778$ nodes and $16,673,476$ edges
and will be used to check the robustness of analyses involving
the distance among users in the Flickr social network.

Similarly, we constructed our Last.fm data set using public API methods (\url{last.fm/api}), in particular for collecting neighbor and friend relations. In Last.fm jargon, friends are contacts in the social networks, while neighbors are users recommended by the system as potential contacts, based on their music playing histories. Last.fm also allows users to annotate various items (songs, artists, or albums) with tags. However, the Last.fm API does not allow to retrieve the complete user annotation activity. Therefore, with permission, we developed a crawler that extracts all the triples (\textit{user, item, tag}) and group membership information by visiting and parsing user profile web pages. The crawls took place over a period of a few months in the first half of 2009. The resulting data set comprises of $99,405$ users, of which $52,452$ are active, i.e., have at least one annotation. The $10,936,545$ triples annotate $1,393,559$ items with $281,818$ tags. The users belong to $66,429$ groups.

No filters were applied to our data set collections.

\section{Data analysis}
\label{sec:analysis}

In this section we present an analysis of the data. The very same analysis 
was carried out for both Flickr and Last.fm data sets. However, due to 
space limitations, we report below 
mainly on the results of the Flickr analysis. Unless otherwise specified our 
analysis refers to $G_0$ but we checked that the results do not change for $G_1$. 
The analysis of Last.fm yielded analogous results, both qualitatively and 
quantitatively. Therefore we believe our conclusions to be robust.

\subsection{Heterogeneity and correlations}
\label{sec:hetero}

\begin{figure}[t]
\centerline{\includegraphics[width=\columnwidth]{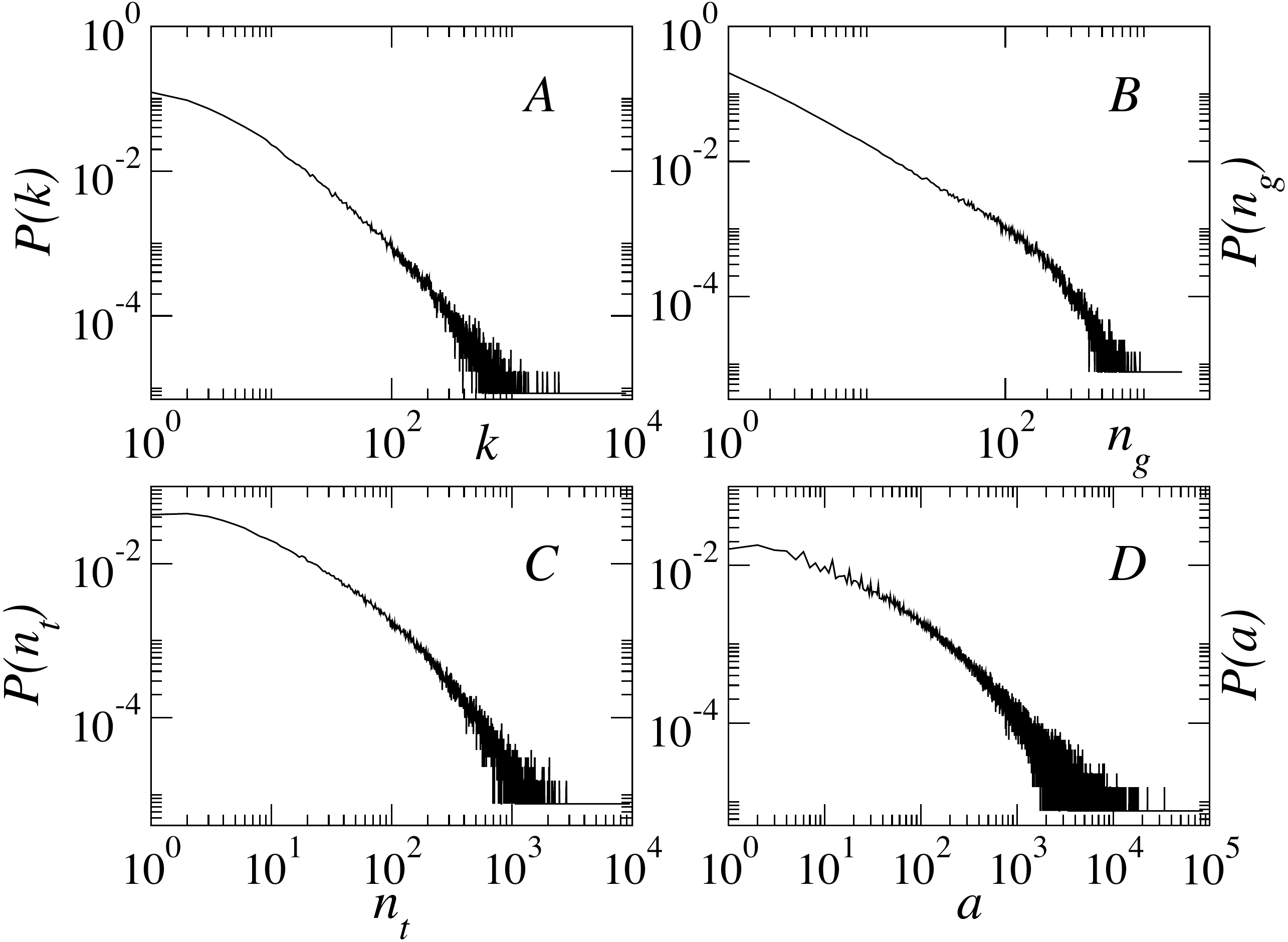}}
\vspace{-1em}
\caption{Flickr distributions of (A) the number $k$ of neighbors of a user,
(B) the number $n_g$ of groups of which a user is a member,
(C) the number $n_t$ of distinct tags per user, and 
(D) the number $a$ of tag assignments per user.}
\label{fig:distr}
\end{figure}

Let us first focus on the activity of users as measured by a number of 
indicators, and investigate the correlations between these indicators.
The activity of a Flickr user has indeed various aspects, among which the most important
are uploading photos, tagging them, participating in groups, commenting on other users'
photos, and other social networking activities. Fig.~\ref{fig:distr} displays the
probability distributions of the number $k$ of neighbors in the social
network (the degree $k$ of a node), and the probability of finding a user
with a given number $n_t$ of distinct tags in her vocabulary.
The breadth of a user's tag vocabulary can be regarded as a proxy
for the breadth of her interests.
We also show in Fig.~\ref{fig:distr} the distribution of
the number of groups $n_g$ to which a user belongs,
and of the total number $a$ of tags assignments submitted by a user
(in this case, a tag used twice by a user is counted twice).
More precisely, if $f_u(t)$ is the number of times that a tag $t$ has been
used by user $u$, then the total number of tag assignments of user $u$ is $a_u =\sum_t
f_u(t)$. All these distributions are broad, showing that the activity
patterns of users are highly heterogeneous. For reference, Table~\ref{table1}
reports the averages and variances of these quantities.

\begin{table}
\centering
\caption{Averages and fluctuations of Flickr user activity}
\begin{tabular}{ccc} \hline
Measure of activity $x$ & Average $\langle x \rangle$ & $\langle x^2 \rangle/\langle x \rangle$  \\ \hline
$k$       &  38.3 &  469 \\ 
$n_{t}$   &  85.3 &  511.4\\ 
$n_{g}$ &  32.6 &  184.6\\ 
$a$ & 690.7 &  8471.3 \\
\hline
\end{tabular}
\label{table1}
\end{table}


A few comments are in order. First, in our analysis we do not
consider one obvious measure of activity, namely the number of photos
uploaded by a user. One reason for this is that the number of photos posted
by a user is known to be strongly correlated with the number of tags 
from the same user~\cite{ht06marlow}. 
More importantly, Flickr is a ``narrow folksonomy'' in which users tend to 
tag mostly their own content~\cite{vanderwal}. Thus, when exploring the similarity of users
and relating it to the underlying social network, shared usage of tags and co-membership
in groups are natural and more direct indicators of shared interests.
%
Another note concerns the comparison with the study by 
Mislove \textit{et al.}~\cite{mislove-2007-socialnetworks},
who reported a smaller average degree for the Flickr social network. 
This difference is due to the fact that in our study we focus on those users who
use both tags and groups. Since only $21\%$ of the users
participate in groups~\cite{mislove-2007-socialnetworks},
this means that here we are focusing on the ``active'' users.
As we will see below, the various activity metrics are correlated with one another,
so users who are more active in terms of tags and groups
will tend to have more contacts in the Flickr social network,
hence the higher average degree we report here.
Fig.~\ref{fig:distr}, however, clearly shows that even within this ``active'' set of users,
very broad distributions of activity patterns are observed
and no ``typical'' value of the activity metrics can be defined.

\begin{figure}[t]
\centerline{\includegraphics[width=0.95\columnwidth]{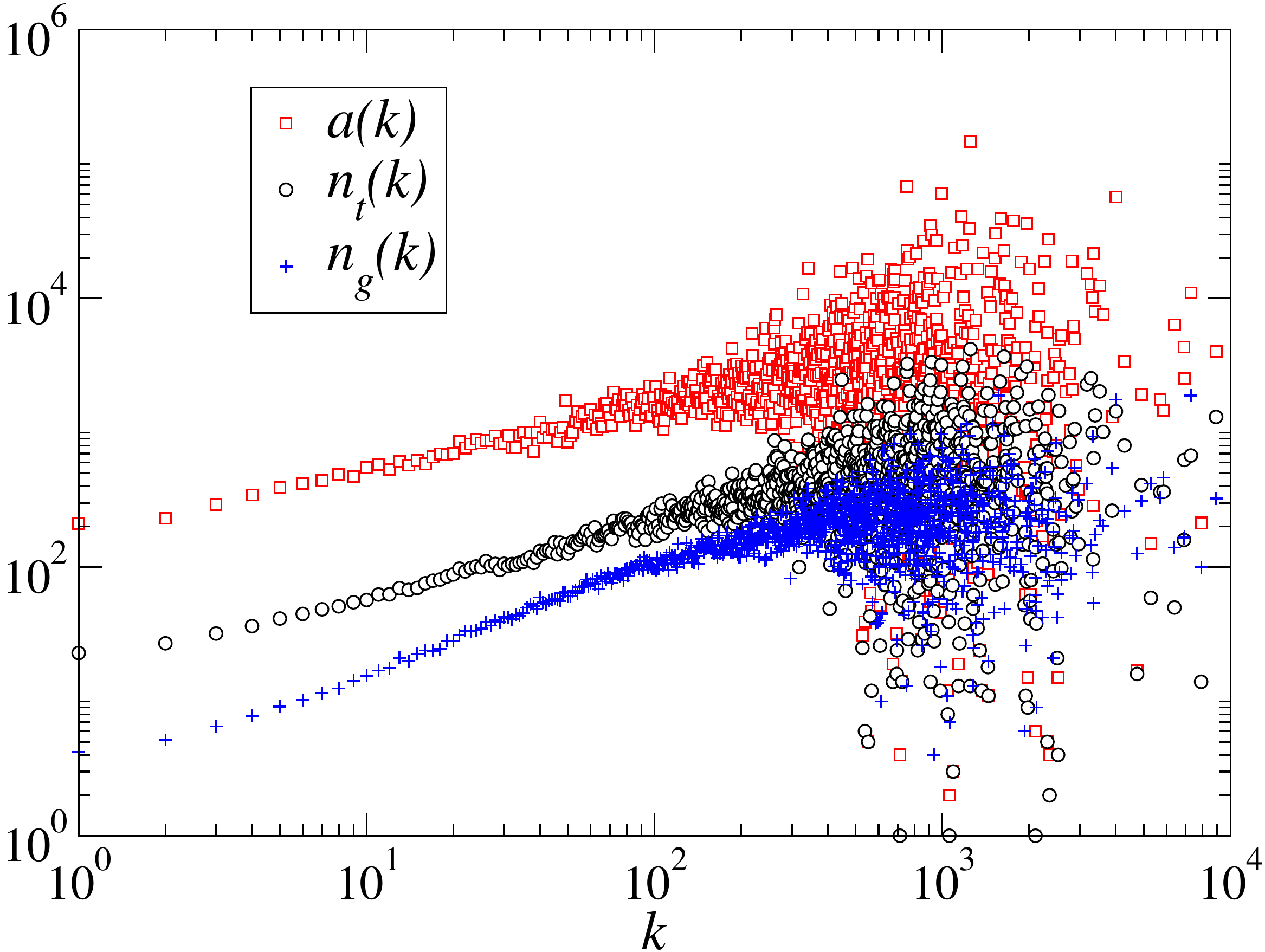}}
\vspace{-1em}
\caption{Average number of distinct tags ($n_t$), of groups ($n_g$), and
of tag assignments ($a$) of users having $k$ neighbors in the Flickr social network.}
\label{fig:correl_k}
\end{figure}

It seems natural to ask whether the different types of activity measures are
correlated with one another and with the structure of the social network:
are users with more social links also more active in
tagging their content, and do they participate in more groups?
As shown in Fig.~\ref{fig:correl_k}, the data show that this is indeed the case (see also Ref.~\cite{ht06marlow}).
Fig.~\ref{fig:correl_k} displays the average activity of users with $k$ neighbors in
the social network, as measured by the various metrics defined above.
For instance,
\[
n_t (k) = \frac{1}{| u: k_u = k |} \sum_{u: \, k_u = k} n_t(u) \, .
\]
All types of activity have an increasing trend for increasing values of $k$,
and of course strong fluctuations are present at all values of $k$.
The strong fluctuations visible for large $k$ values are due to the fewer
highly-connected nodes over which the averages are performed.
Notably, users with a large number of social contacts but using very few tags
and belonging to very few groups can be observed. Despite these important heterogeneities
in the behavior of users with the same degree $k$, the data clearly indicate a strong
correlation between the different types of activity metrics. The Pearson correlation
coefficients are: 0.349 between $k$ and $n_t$, 0.482 between $k$ and $n_g$,
0.268 between $k$ and $a$, 0.429 between $n_t$ and $n_g$, 0.753 between $n_t$ and $a$,
0.304 between $n_g$ and $a$.

Another important question concerns the correlations
between the activity metrics of users who are linked in the
Flickr social network. This is a well-known problem in the 
social sciences, ecology, and epidemiology:
a typical pattern, referred to as ``assortative mixing,'' describes the tendency of nodes
of a network (here, the users) to link to other nodes with similar properties.
This appears intuitive in  the context of a social network, where one expects individuals
to connect preferentially with other individuals sharing the same interests.
Likewise, it is possible to define a ``disassortative mixing'' pattern
if the elements of the network tend to link to nodes that
have different properties or attributes.
Mixing patterns can be defined with respect to any
property of the nodes~\cite{Newman:2003d}. In the present case,
we characterize the mixing patterns concerning the various activity metrics.

In the case of large scale networks, the
most commonly investigated mixing pattern involves the degree (number of neighbors) of nodes.
This type of mixing refers to the likelihood that users
with a given number of neighbors connect with users with similar degree.
To this end, a commonly used measure is given by the average nearest neighbors degree of a user $u$,
\[
k_{nn}^u = \frac{1}{k_u}\sum_{v \in {\cal V}(u)} k_v \ ,
\]
where the sum runs over the set ${\cal V}(u)$ of nearest neighbors of $u$.
To characterize mixing patterns in the degree of nodes, a convenient measure
can be built on top of $k_{nn}^u$ by averaging over all nodes $u$ that have
a given degree $k$~\cite{Pastor:2001c,Vazquez:2002b}:
\begin{equation}
k_{nn}(k)=\frac{1}{|u: k_u=k|} \sum_{u: \, k_u=k} k_{nn}^u \ .
\label{eq:knn}
\end{equation}

In the case of Flickr, each user is endowed with several properties characterizing
its activity. It is thus interesting to characterize the mixing  patterns with respect to
all of those properties. To this end, we generalize the average nearest neighbors degree
presented above, and define for each user $u$ the following quantities:
(i) the average number of tags of its nearest neighbors,
\[
n_{t,nn}^u= \frac{1}{k_u}\sum_{v \in {\cal V}(u)} n_{t}(v) \ ,
\]
(ii) the average total number of tags used by its nearest neighbors,
\[
a_{nn}^u= \frac{1}{k_u}\sum_{v \in {\cal V}(u)} a(v) \ ,
\]
and (iii) the average  number of groups to which its nearest neighbors participate,
\[
n_{g,nn}^u = \frac{1}{k_u}\sum_{v \in {\cal V}(u)} n_{g}(v) \ .
\]

To detect the mixing patterns, in complete analogy with the case of
$k_{nn}(k)$, we compute the average number of distinct tags of the nearest neighbors
{\em for the class of users having $n_t$ distinct tags}:
\begin{equation}
n_{t,nn}(n)=\frac{1}{|u: n_{t}(u)=n|} \sum_{u: n_{t}(u)=n} n_{t,nn}^u \ ,
\label{eq:tnn}
\end{equation}
the average total number of tags used by the nearest neighbors
{\em for the class of users with $a$ tag assignments}:
\begin{equation}
a_{nn}(a)= \frac{1}{|u: a(u)=a|}\sum_{u: a(u)=a} a_{nn}^u \ ,
\label{eq:ann}
\end{equation}   
and the average number of groups of the nearest neighbors
{\em for the class of users who are members of $n_g$ groups}:
\begin{equation}
n_{g,nn}(n)=\frac{1}{|u: n_{g}(u)=n|} \sum_{u: n_{g}(u)=n} n_{g,nn}^u \ .
\label{eq:gnn}
\end{equation}

\begin{figure}[t]
\centerline{\includegraphics[width=\columnwidth]{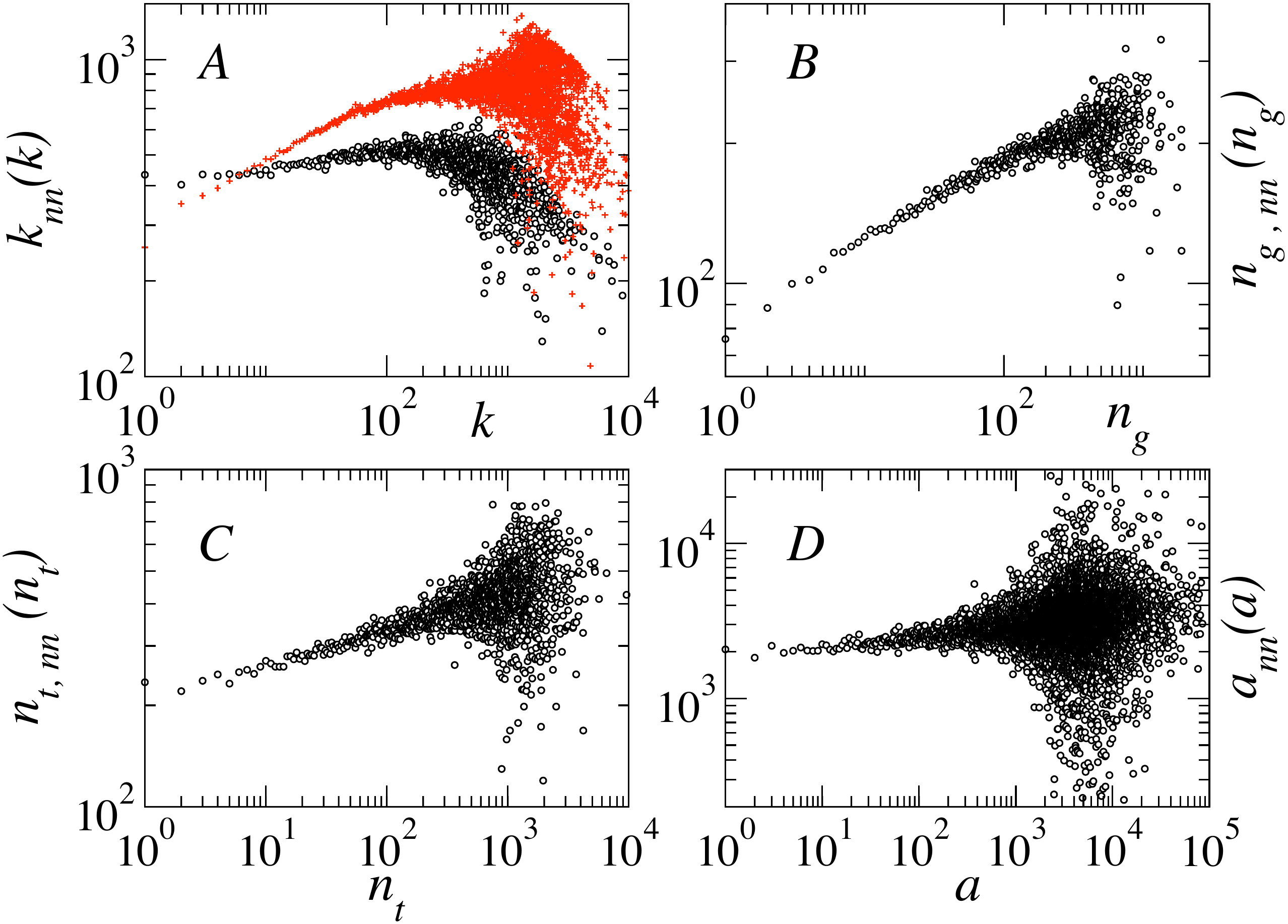}}
\vspace{-1em}
\caption{(A) Average degree of the nearest neighbors of nodes
of degree $k$, computed for $G_0$ (black circles) $G_1$ (red crosses) Flickr networks.
(B) Average number of groups for the nearest neighbors of nodes belonging to $n_{g}$ groups.
(C) Average number of tags for the nearest neighbors of nodes with $n_{t}$ tags.
(D) Average total number of tag assignments for the nearest neighbors of nodes with $a$ tag assignments. In all cases a clear assortative trend is observed.}
\label{fig:assortative}
\end{figure}

Fig.~\ref{fig:assortative} displays the quantities of
Eqs.~\ref{eq:knn}--\ref{eq:gnn}
for the Flickr data set. In all cases, a clear assortative trend is visible:
the average activity of the neighbors of a user increases with the user's
own activity, for all the activity measures we computed. Note that for
the degree mixing patterns, the assortative trend is even enhanced when $G_1$ is
considered instead of $G_0$.
Large fluctuations are observed for large activity values,
because of the small number of very active users.
We remark that while Mislove \textit{et al.}~\cite{mislove-2007-socialnetworks}
had already found an assortative trend with respect to the degree of the social network,
Fig.~\ref{fig:assortative} highlights that the activities of socially connected
users are correlated at all levels.

%

\subsection{Lexical and topical alignment}
\label{sec:alignment}

In this section we analyze more in detail the similarity of user profiles
in relation to their social distance.
More precisely, the previous section was devoted to the correlations between
the \textit{intensity} of user activities, as quantified by several metrics.
We now focus on the \textit{similarity} between user profiles
as measured by the similarity of their respective tag vocabularies,
and by the similarity of the set of groups they belong to.

As mentioned above, Flickr is a ``narrow folksonomy''~\cite{vanderwal}:
tag annotations are provided mostly by the content creator,
i.e., the tags associated with a photo are typically provided by the user
who posted that photo. Intuitively, the absence of shared content,
together with the very personal character of both the content and the tag metadata,
make the Flickr tag vocabulary extremely incoherent across the user community.
Conversely, social bookmaking systems like Delicious allow multiple users
to annotate the same resource and one could argue that the browsing experience
exposes users to the global tag vocabulary and fosters ---~at least in principle~---
imitative or cooperative processes leading to the emergence of global conventions
in the user community~\cite{mathes04folksonomies}.

In light of the above observations, we do not expect to observe
a globally shared tag vocabulary in the Flickr community.
A simple test for the existence of such a globally shared vocabulary
can be performed by selecting pairs of users at random
and measuring the number of tags they share, $n_{st}$.
It turns out that the average number of shared tags is only 
$\langle n_{st} \rangle \approx 1.6$.
The most common case (mode) is in fact the absence of any shared tags;
this occurs with probability close to $2/3$ among randomly chosen pairs of users.


One can nevertheless expect that a number of mechanisms may lead to \textit{local} alignment
of the user profiles, in terms of shared tags and/or groups membership.
The presence of a social link, in fact, indicates a priori some degree of shared context
between the connected users, which are likely to have some interests
in common, or to share some experiences, or who are simply exposed
to each other's content and annotations.
As an example, Table~\ref{table2} shows the $12$ most frequently used tags
for three Flickr users with comparable tagging activity.
User~\textit{A} and user~\textit{B} have marked each other as friends,
while user~\textit{C} has no connections to either \textit{A} or \textit{B}
on the Flickr social network. All of these users have globally popular
tags in their tag vocabulary. In this example, the neighbors \textit{A} and \textit{B} share an interest (expressed by the tag \textit{flower}) and several of the most frequently used tags (marked in bold).

Regardless of the mechanism driving
this potential local alignment, in the following we want to measure this effect for
the case of tag dictionaries and group memberships, and put it into relation
with the distances between users along the social network. 
This approach is similar to the exploration of topical locality in the Web, 
where the question is whether pages that are closer to 
each other in the link graph are more likely to be related to one 
another~\cite{Menczer01topo}. 

\begin{table}
\centering
\caption{Tags most frequently used by three Flickr users}
\begin{tabular}{ccc} \hline
User~\textit{A}        & User~\textit{B}    & User~\textit{C} \\ \hline
\textbf{green}         & \textbf{flower}    & japan        \\ 
\textbf{red}           & \textbf{green}     & tokyo        \\ 
catchycolors           & kitchen            & architecture \\ 
\textbf{flower}        & \textbf{red}       & bw           \\ 
\textbf{blue}          & \textbf{blue}      & setagaya     \\ 
\textbf{yellow}           & white              & reject       \\ 
catchcolors            & fave                & sunset       \\ 
travel                 & detail             & subway       \\ 
london                 & closeupfilter      & steel        \\ 
pink                   & metal              & geometry     \\ 
orange                 & \textbf{yellow}    & foundart     \\ 
macro                  & zoo                & canvas       \\ \hline
\end{tabular}
\label{table2}
\end{table}

First, we must define robust measures
of vocabulary similarity and group affiliation similarity between two users $u$
and $v$. The first and simplest measures are given by the number of
shared tags $n_{st}$ among the tag vocabularies of $u$ and $v$, and by the number of
shared groups $n_{sg}$ to which both $u$ and $v$ belong.
These measures, however, are not normalized and can be affected
by the specific activity patterns of the users: two very active users may have more tags
in common than two less active users, just because active users tag more, on average.
We therefore consider as well a distributional notion of similarity between the
tag vocabularies of $u$ and $v$. Following Ref.~\cite{cattuto2008iswc}
we regard the vocabulary of a user as a feature vector whose elements
correspond to tags and whose entries are the tag frequencies
for that specific user's vocabulary. To compare the tag feature vectors
of two users, we use the standard cosine similarity~\cite{salton89}.
Denoting by $f_u(t)$ the number of times that tag $t$ has been used
by user $u$, the cosine similarity $\sigma_{tags}(u,v)$
is defined as
\begin{equation}
\sigma_{tags}(u,v) = \frac{\sum_t f_u(t) f_v(t) }{ \sqrt{\sum_t f_u(t)^2} \sqrt{\sum_t f_v(t)^2} } \, .
\end{equation}
This quantity is $0$ if $u$ and $v$ have no shared tags, and $1$ if
they have used exactly the same tags, in the same relative proportions.
Because of the normalization factors in the denominator,
$\sigma_{tags}(u,v)$ is not directly influenced by the global activity
of a user.

Similarly, we can define the cosine similarity $\sigma_{groups}$
for groups memberships. Since a user belongs at most once to a
group, this reduces to
\begin{equation}
\sigma_{groups}(u,v) =
\frac{\sum_g \delta_u^g \delta_v^g }{ \sqrt{n_{g}(u) n_{g}(v)} }
\end{equation}
where $\delta_u^g = 1$ if $u$ belongs to group $g$ and $0$ otherwise.

To compute averages of these similarities, we randomly chose $N=2 \times 10^4$ users
and explored their neighborhoods in a breadth-first fashion.
In order to exclude biases due to this sampling, we also performed
an exhaustive investigation of the social network neighborhoods up to distance $2$ from each user,
obtaining the same results. Moreover we considered the distances along
the social network using $G_1$ instead of $G_0$, and again found the same results, showing
the robustness of the observed behavior with respect to possible sampling biases due
either to the crawl or to considering only users having both tagging
activity and groups memberships.


\begin{figure}[t]
\centerline{\includegraphics[width=\columnwidth]{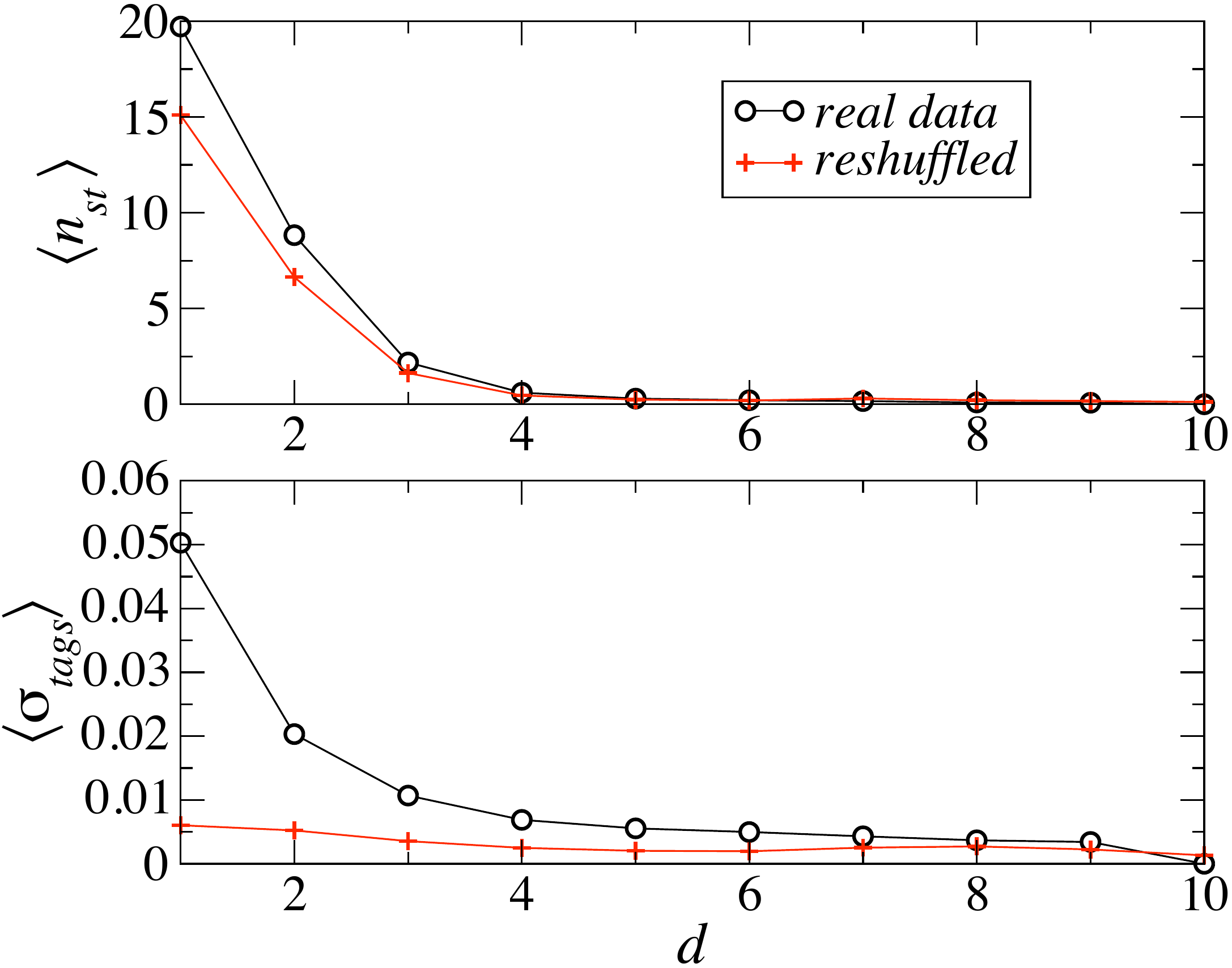}}
\vspace{-1em}
\caption{Top: average number of shared tags $\langle n_{st}\rangle$
  for two Flickr users as a function of their distance $d$ along the social
  network. Bottom: average cosine similarity $\langle \sigma_{tags}
  \rangle$ between the tag vocabularies of two Flickr users as a function of
  $d$. In both cases data for the same social network with reshuffled
  tag vocabularies
are shown.}
\label{fig:ovtags_vs_d}
\end{figure}

\begin{figure}[t]
\centerline{\includegraphics[width=\columnwidth]{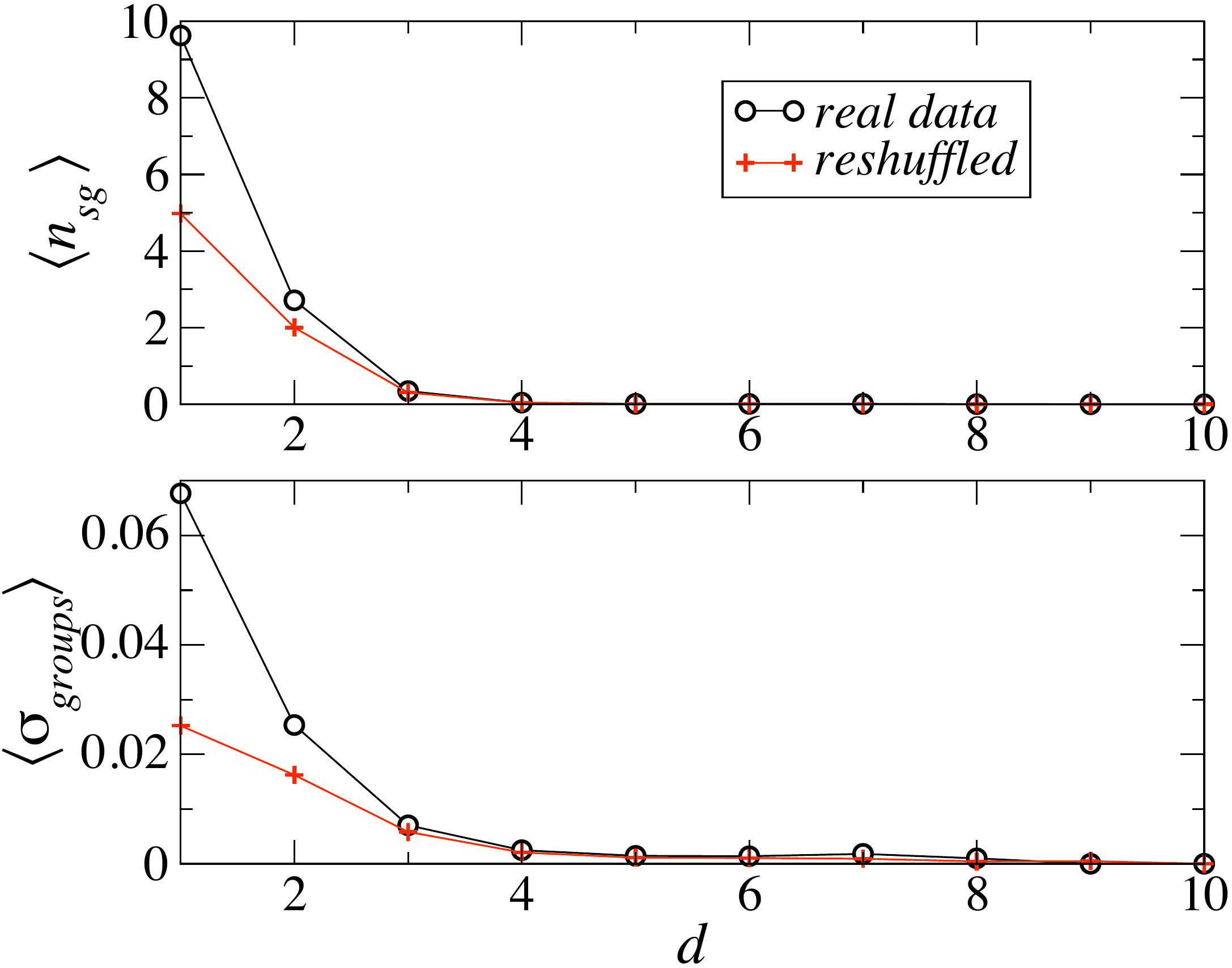}}
\vspace{-1em}
\caption{Top: average number of shared groups $\langle n_{sg}\rangle$
  for two Flickr users as a function of their distance $d$ along the social
  network. Bottom: average cosine similarity of the group
  affiliation $\langle \sigma_{groups} \rangle$ vs $d$.  In both cases
  data for the same social network with reshuffled group affiliations
  (preserving the number of groups for each user) are shown.}
\label{fig:ovgroups_vs_d}
\end{figure}

Figures~\ref{fig:ovtags_vs_d} and \ref{fig:ovgroups_vs_d} give an
indication of how the similarity between users depends
on their distance $d$  along the social network, by showing the
average number of shared tags, of shared groups, and the
corresponding average cosine similarities, of two users
as a function of $d$. While the average number of shared tags or groups
is quite large for neighbors (respectively close to $20$ and to $10$), it
drops rapidly (exponentially) as $d$ increases, and is close to $0$ for $d \geq 4$.

Figures~\ref{fig:ovtags_vs_d} and \ref{fig:ovgroups_vs_d}
provide a strong indication that a certain
degree of alignment between neighbors in the social network is
observed both at the lexical level and for group
affiliations. As soon as the distance between two users on the social
distance is not $1$ (neighbors) or $2$ (neighbors
of neighbors) however, it becomes highly probable that these users
have neither tags nor groups in common. Therefore the alignment is
a strongly local effect.

The analysis of the mixing patterns of the social network
performed previously leads us to investigate in more detail
this local alignment. This analysis has indeed shown the presence of a strong
assortativity with respect to the intensity of the users' activity.
It could therefore be the case that such assortativity,
by a purely statistical effect, yields an ``apparent'' local alignment
between the tag vocabularies of users. For example, even in a hypothetical
case of purely random tag assignments, it would be more probable to
find tags in common between two large tag vocabularies than between a small one
and a large one.

In order to discriminate between effects due to actual lexical and group
membership similarity and those simply due
to the assortativity, it is important to devise a proper \textit{null model},
i.e. to construct an artificial system that retains the same social structure
as the one under study, but lacks any lexical or topical alignment other than the
one that may result from statistical effects. This
is done by keeping fixed the Flickr social network and its assortativity pattern
for the intensity of activity, but destroying socially-related lexical or topical alignments
by means of a random permutation of tags among themselves
and groups among themselves. More precisely, we proceed in the following fashion:
(i) we keep the social network unchanged;
(ii) we build the global list of tags with their multiplicity,
i.e. each tag appears the total number of times it has been used;
(iii) for each user with $n_t$ tags $t_1, t_2 \ldots t_{n_t}$,
with respective frequencies $f_1, f_2, \ldots, f_{n_t}$,
we extract $n_t$ distinct tags at random from the global list of tags
and assign them to $u$ with frequencies $f_1, f_2, \ldots, f_{n_t}$.
This guarantees that the number of distinct tags and the total
number of tag assignments for each user is the same as in the original data, and
that the distribution of frequencies of tags is left unchanged.
Clearly, this null model preserves the assortativity patterns with respect to
the amount of activity of users, as each user has exactly the same
number of distinct tags and of tag assignments as in the real data. However, correlations
between the tag vocabularies are lost, except the ones
purely ascribed to statistical effects.

For group membership, we can proceed in a similar way:
(i) we build a list of groups with a multiplicity equal to the number
of users of each group (i.e., a group appears $n$ times in the list if
it has $n$ users);
(ii) for each user $u$ belonging to $n_g$ groups,
we extract at random $n_g$ (distinct) groups from the list and assign
them to $u$. As for the tags, this procedure preserves the number of groups for each
user, as well as the statistics of the number of users per group, while destroying
correlations between users' group memberships.

The goal of the null model is to determine the amount of  
lexical and topical alignment due to spurious activity correlations. 
Eliminating such spurious correlations is analogous in purpose to the  
use of inverse document frequency (IDF) in information retrieval. 
IDF discounts the contribution of globally common terms in assessing 
the similarity between documents. Such terms are likely to 
be shared by pairs of documents solely because of their statistical 
prevalence. Unlike in information retrieval, it is not straightforward 
to apply this type of discounting in social annotations. One would first 
need to determine whether to discount tags based on their prevalence 
among users or among resources. The null model destroys all spurious 
correlations regardless of their source.

Using the null model, we measured the alignment between users at
distance $d$ on the social network in the same way as for the
original data. As Figs.~\ref{fig:ovtags_vs_d} (top) and
\ref{fig:ovgroups_vs_d} (top) show, the average number of shared tags
or of shared groups, as a function of the distance $d$, shows a 
similar trend to the original (non-reshuffled) data. For neighbors
and next to nearest neighbors ($d<3$), the average numbers of
shared tags or groups are 
lower in the null model than in the original data, but still 
significantly higher than for users at larger distances. 
The assortative mixing between the 
amount of activity of neighboring users is therefore enough to yield
a strong lexical and topical alignment {\em as simply measured by the
number of shared tags or groups}. The case of cosine similarity is
quite different. As shown in Fig.~\ref{fig:ovtags_vs_d} (bottom),
the average cosine similarity is very small in the null model, and
does not depend strongly on the distance in the social network. 
Therefore local lexical alignment is a real effect: friends are 
more likely to use similar tag patterns. 
With respect to the topical alignment, a certain --- albeit weaker --- 
dependence of $\langle \sigma_{groups} \rangle$ on $d$ is visible
in Fig.~\ref{fig:ovgroups_vs_d} (bottom). 

\begin{figure}[t]
\centerline{\includegraphics[width=\columnwidth]{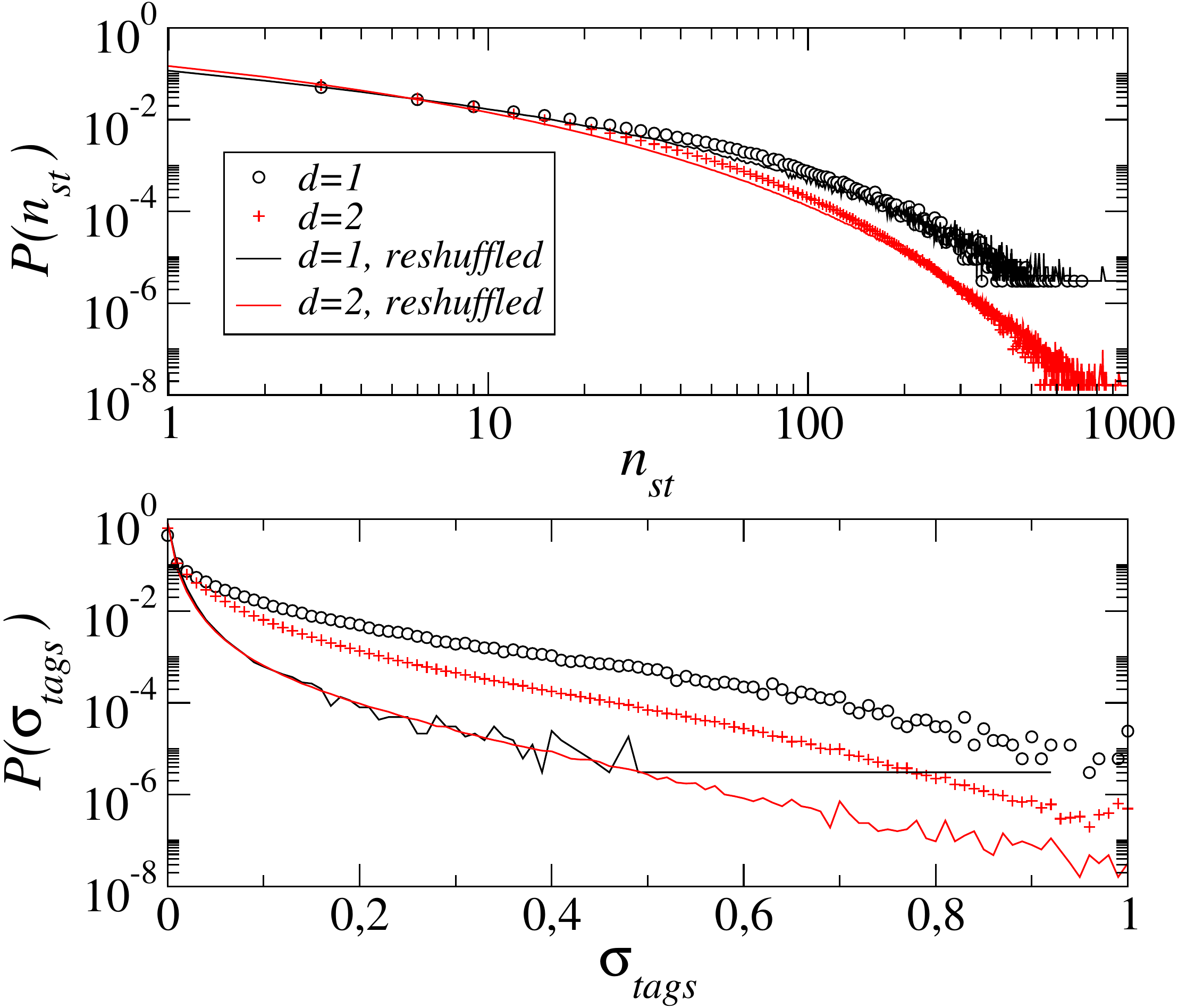}}
\vspace{-1em}
\caption{Top: Probability distributions of the number of shared tags
  of two users being at distance $d$ on the social network, for $d=1$
  and $d=2$ (symbols), and for the same network with reshuffled tags
  (lines). Bottom: same for the distributions of cosine
  similarities of the tag vocabularies.}
\label{fig:prob_ov_null}
\end{figure}

We also analyzed the distributions of $n_{st}$, $\sigma_{tags}$, 
$n_{sg}$, and $\sigma_{groups}$ among users at fixed distance $d$,
for both the original and the reshuffled data.
For brevity we show only the distributions of $n_{st}$ and $\sigma_{tags}$
for $d=1$ and $d=2$
in Fig.~\ref{fig:prob_ov_null}. The distributions of $n_{st}$ 
are very similar for the original and the reshuffled data, while 
for the cosine similarity they are clearly different: 
a much stronger local alignment occurs in the original data.

\begin{figure}[t]
\centerline{\includegraphics[width=\columnwidth]{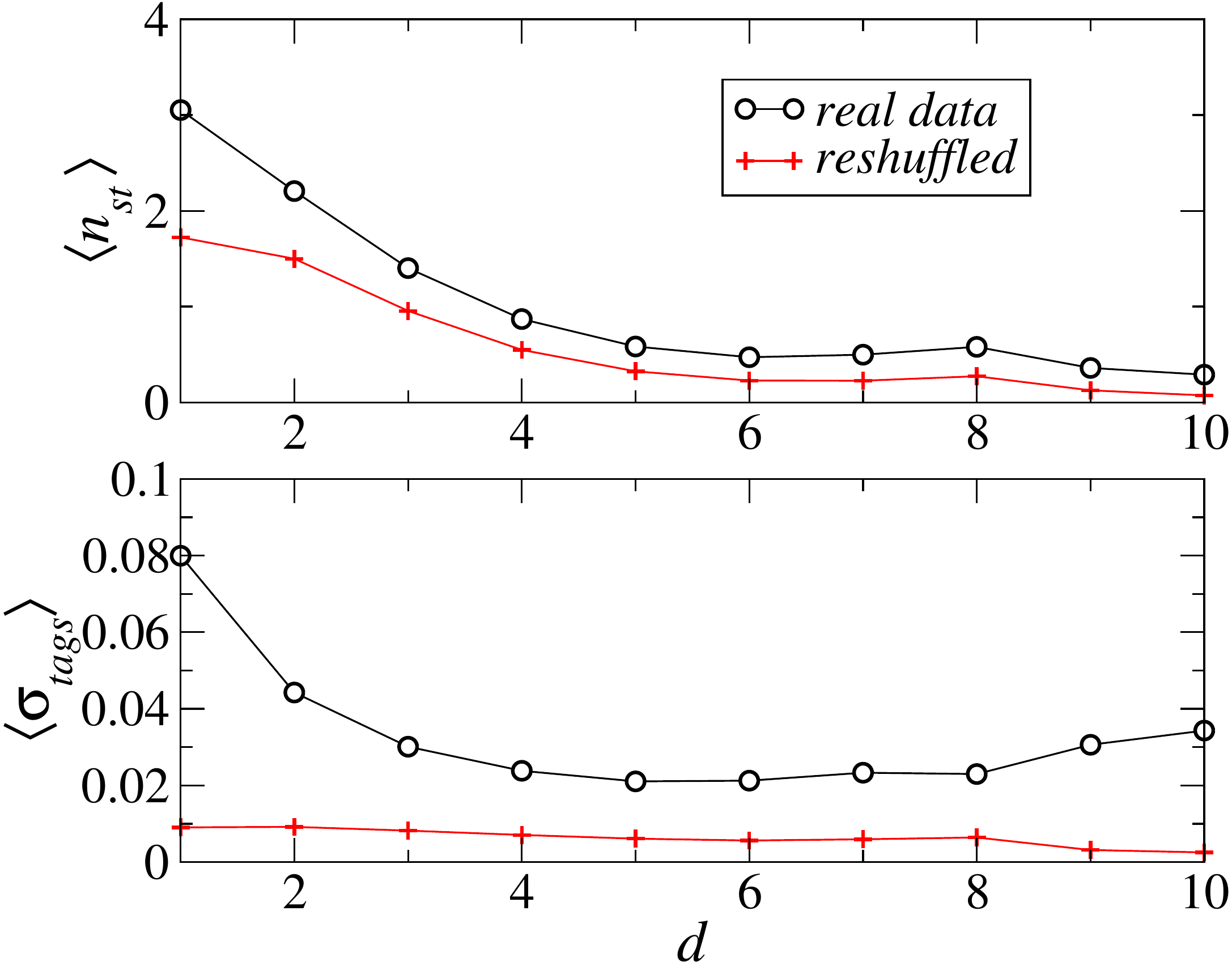}}
\vspace{-1em}
\caption{Average number of shared tags (top) and average cosine similarity 
between tag vocabularies (bottom) for pairs of Last.fm users as a 
function of their social distance. We also show data for the same social 
network with reshuffled tag vocabularies.}
\label{fig:ovtags_vs_d_lastfm}
\end{figure}

As mentioned earlier, analogous results are found by analyzing our 
Last.fm data set. For illustration purposes we just show in 
Fig.~\ref{fig:ovtags_vs_d_lastfm} the dependencies of local 
tag alignment measures on social distance. Again cosine similarity 
is the more robust measure.

Our investigation of the lexical and topical alignment patterns
in Flickr and Last.fm reveals therefore the following picture. 
The various measures of the topical and lexical overlap between users
as a function of their distance along the social network clearly point
toward a partial local alignment, which persists up to distances $2-3$, even if
large values can occasionally still be observed at larger distances. Interestingly,
if the number of shared tags between users is the only retained measure,
a reshuffling of tags and groups between users shows that a large part
of the alignment is simply due to the assortative pattern concerning users'
amounts of activity. This result highlights the importance of considering
appropriate null models to discriminate between purely statistical
effects and real lexical or topical alignments. It also shows that
correctly normalized similarity measures such as cosine similarity, which
factor out the effects of vocabulary size, are more appropriate for
such investigations, since they are less affected by the assortativity
patterns.

%


\section{Predicting social links}
\label{sec:predict}

The analysis in the previous section strongly suggests that users with 
similar topical interests, as captured by shared tags in particular, are more 
likely to be neighbors in the social network. Therefore a natural question 
is whether semantic similarity measures among users based solely on their 
annotation patterns can be employed as accurate predictors of friendship links. 
We tested this hypothesis on both our Flickr and Last.fm data sets, 
because each provides annotation metadata needed to compute similarity as well as 
a social network to evaluate the accuracy of the predictions. 
For brevity we focus on reporting the results for the Last.fm data, which 
are more interesting for two reasons.  First, contrary to Flickr, Last.fm 
is a ``broad folksonomy'' in which different users can easily annotate the same  
songs, artists, or albums. This allows us to compute similarity based on 
shared content as well as shared vocabulary. Second, Last.fm provides 
\emph{neighbor} recommendations. Neighbors are users with a similar music taste, 
based on listening patterns. The neighborhood relation is therefore 
independent of the explicit friendships established by the users, and 
provides an obvious gauge against which to evaluate any algorithm to predict 
social links. Except for the lack of such a comparison measure in Flickr 
(beyond the random choice baseline), and for not considering 
similarity measures based on shared items in Flickr, the prediction analysis 
yields consistent and encouraging results using both data sets. 

\subsection{Overview of semantic similarity measures}

In prior work~\cite{Markines08HT, Markines09www, ht09_GaL_MIP_poster} 
we evaluated a number of \emph{social similarity} measures based on 
folksonomies, i.e., on annotations represented as \emph{triples} 
(\textit{user, item, tag}) where Flickr photos and Last.fm songs are
instances of items. All of these social similarity measures have the 
desirable property of being symmetric in the sense that they can be 
directly applied to compute the similarity between two items, two 
tags, or two users from a folksonomy. Therefore we employ several of 
these measures here to predict social network 
links from the similarity among users. We summarize below a few main 
features of the proposed user similarity measures; for further details 
and examples see Refs.~\cite{Markines09www, ht09_GaL_MIP_poster}.

We consider two aggregation schemes. In \emph{distributional} aggregation, 
we project along one of the dimensions keeping track of frequencies. 
For example, projecting onto items, a user $u$ is represented as a tag vector
whose component $f_u(t)$ is the number of items tagged by $u$ with $t$. 
Analogously we can project onto tags representing users as item vectors.  
Unfortunately distributional aggregation requires 
that all similarities be recomputed for any change in annotations, 
leading to quadratic runtime complexity.

In \emph{collaborative} aggregation, first we pick a feature (tag or item)
and for each value of this feature we represent each user as a list of 
values of the other feature (items or tags). Then we compute a different 
similarity value between two users according to each of these lists. 
Finally we aggregate these similarities by voting (summing). For example, 
for each tag we can compute a similarity value based on item lists. 
These are then summed across tags to obtain the final similarity.
Analogously we can compute similarities from tag lists and sum 
them across items. Collaborative aggregation has two advantages. 
First, it can be integrated with collaborative filtering techniques 
(hence the name) by a judicious definition of conditional probabilities 
$p(item|tag)$ or $p(tag|item)$. This makes collaborative similarity 
measures competitive with distributional measures in terms of accuracy~\cite{Markines09www}.  
Second, similarities based on collaborative aggregation can be 
updated incrementally, in linear time. When a triple is added or deleted, 
only similarities involving the item or tag in that triple need be 
updated. As a result, collaborative aggregation leads to \emph{scalable} 
social similarity measures. 
Each aggregation scheme has two variants, depending on whether we project 
onto/aggregate across tags or items.

For each aggregation scheme/variant we consider six measures: 
\emph{cosine, overlap, matching, Dice and Jaccard coefficients, 
and maximum information path (MIP).} 
Note that distributional cosine with projection onto tag 
vectors is the $\sigma_{tags}$ measure discussed in the previous section. 
MIP is a generalization of Lin's similarity~\cite{lin98information} 
to the non-hierarchical triple representation~\cite{ht09_GaL_MIP_poster}. 
For example, the distributional version of MIP with aggregation across 
items is defined as 
\[
\sigma^{MIP}_{items}(u_1,u_2) = \frac{
	2 \log(min_{t \in T_1 \cap T_2} p[t])
}{
	\log(\min_{t \in T_{1}} p[t]) + 
	\log(\min_{t \in T_{2}} p[t])
}
\] 
where $T_i$ is the set of tags used by $u_i$ and $p[t]$ is the fraction of users annotating with tag $t$. For aggregation across tags the definition is analogous except that we look at probabilities of shared items. 
For the collaborative version projecting onto items, say, we would similarly define $\sigma^{MIP}_{items}(u_1,u_2; r)$ for each item $r$ replacing $T_i$ by the set $T^r_i$ of tags used by $u_i$ to annotate $r$, and replacing $p[t]$ by a suitably defined $p[t|r]$. Finally $\sigma^{MIP}_{items}(u_1,u_2) = \sum_r \sigma^{MIP}_{items}(u_1,u_2; r)$. 
Among the measures discussed in Ref.~\cite{Markines09www} we did not consider 
mutual information due to its higher computational complexity. In addition 
to these $6 \times 2 \times 2 = 24$ measures, we also consider for comparison 
purposes the affinity score provided by Last.fm for the 60 top neighbors of each user. As mentioned earlier, this score is based on similar music taste and computed from listening patterns.

\subsection{Methodology}

The evaluation consists in selecting a set of pairs of users, computing each similarity measure for each pair, and adding social links between users in decreasing order of their topical similarity: the pairs of users with highest similarity are those we predict to be most likely friends. For each predicted social link, we check the actual social network to see if the prediction is correct. As one decreases the similarity threshold more links are added, leading to more true positives but also more false positives. The best similarity measure is the one that achieves the best ratio of true positive to false positive rate across similarity values, as illustrated by ROC plots and quantified by the area under the ROC curve (AUC).

To sample the pairs of users from our data set, we start by sorting the users by one of three different criteria:
\begin{enumerate*} 
\item{\emph{Most Active:}} By number of annotations;
\item{\emph{Most Connected:}} By number of friends;
\item{\emph{Random:}} By shuffling.
\end{enumerate*} 
The set $P$ of pairs is then constructed according to the following algorithm:
\begin{alltt}
  repeat:
    pick next \(u\) by sorting criterion
    \(R \leftarrow\) set of 60 neighbors of \(u\)
    for each \(n\) from \(R\):
      if \(n\) is active:
        \(P \leftarrow (u,n)\)
        stop when \(|P| = M\)
\end{alltt}
Recall that users are considered active if they have at least one annotation. This is a requirement in order to compute topical similarity. The choice to select pairs among neighbors stems from the goal of comparing the accuracy of topical similarity methods with Last.fm recommendations. Given the sparsity of the social and neighbor networks, comparative evaluation would be impossible without such a sampling. Note that this sampling strategy may bias the evaluation in favor of Last.fm's neighbor recommendations, because if two active neighbors are friends, they are guaranteed to be detected while two active friends who are not neighbors would be missed by our sampling even if they were detected by our similarity measures. Therefore our sampling algorithm is a conservative choice in that it does not unfairly help our similarity measures in the evaluation.\footnote{At press time Last.fm has released a new API functionality, called \emph{Tasteometer}, to query the affinity score for arbitrary user pairs. This will allow us to sample users independently of neighborhood relations in future evaluations.}   
We experimented with sets of pairs of cardinality $M=1,000$ and $M=2,500$. The results are similar; we report below on evaluations with 1,000 pairs.

\newpage

\subsection{Results}

\begin{figure}
\centering
\begin{tabular}{cc}
\includegraphics[width=0.47\columnwidth]{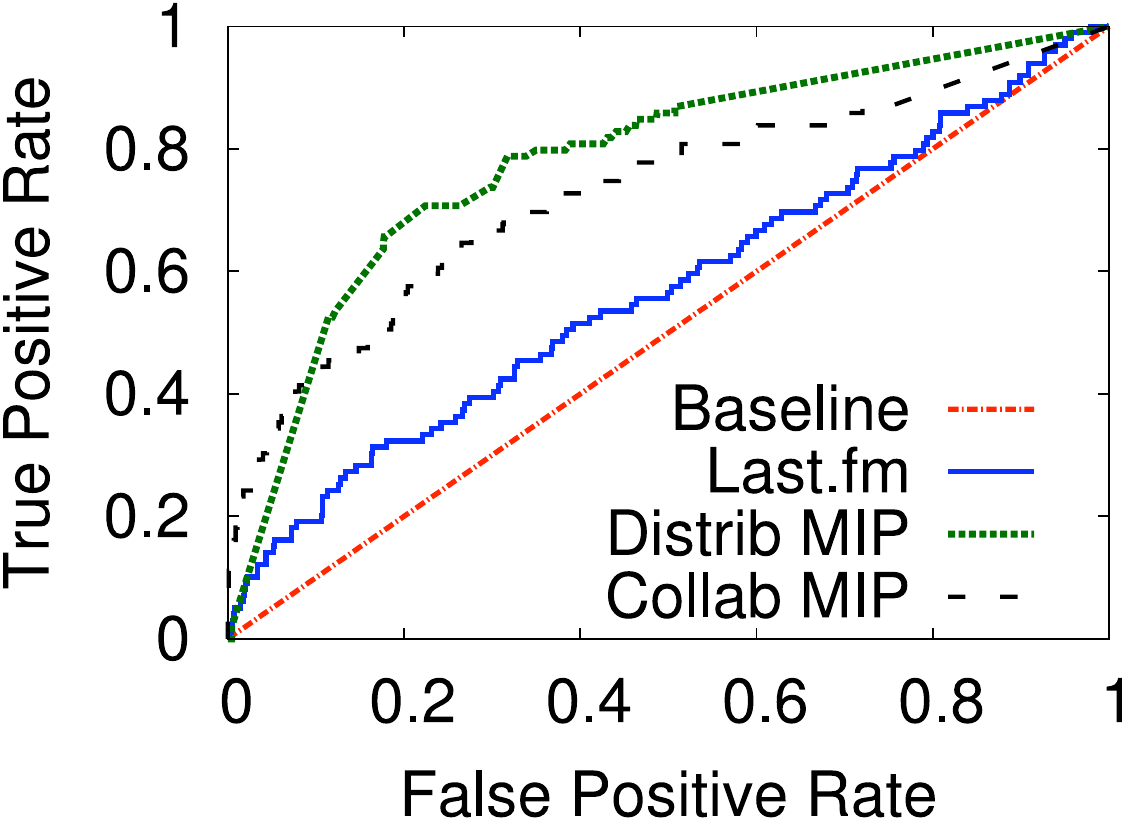} &
\includegraphics[width=0.47\columnwidth]{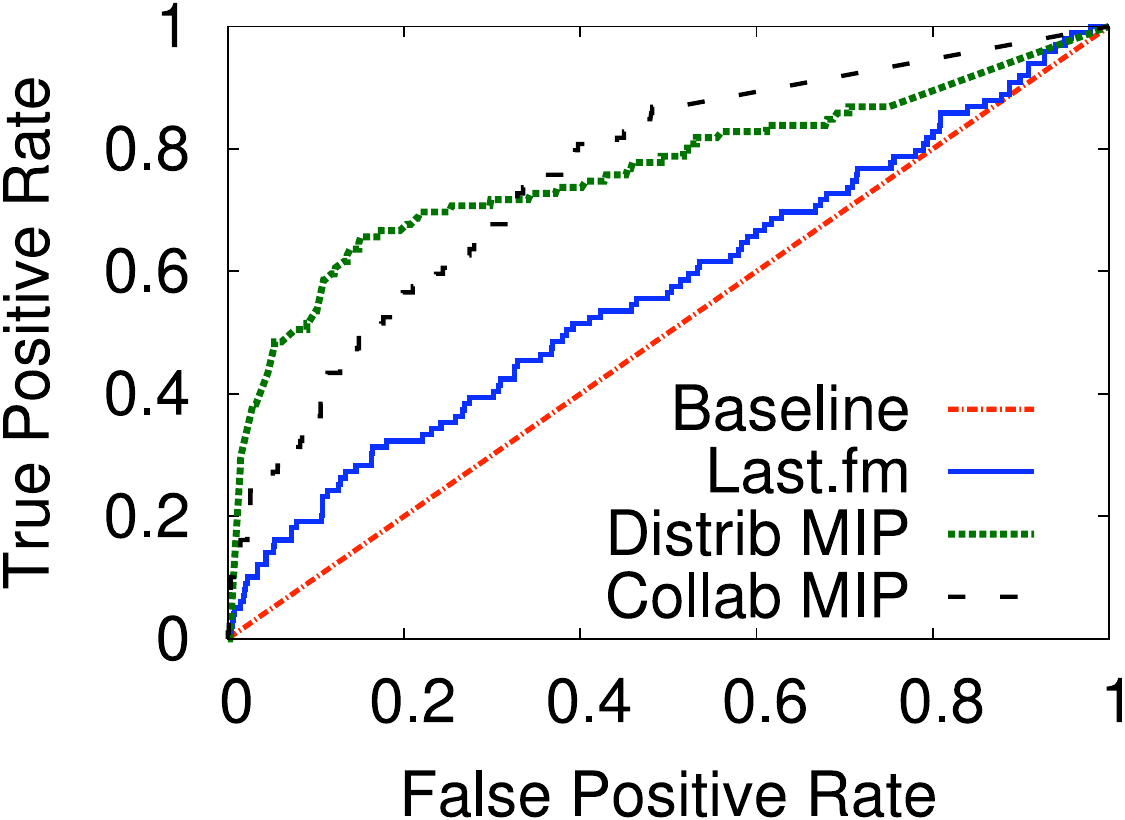} \\
\includegraphics[width=0.47\columnwidth]{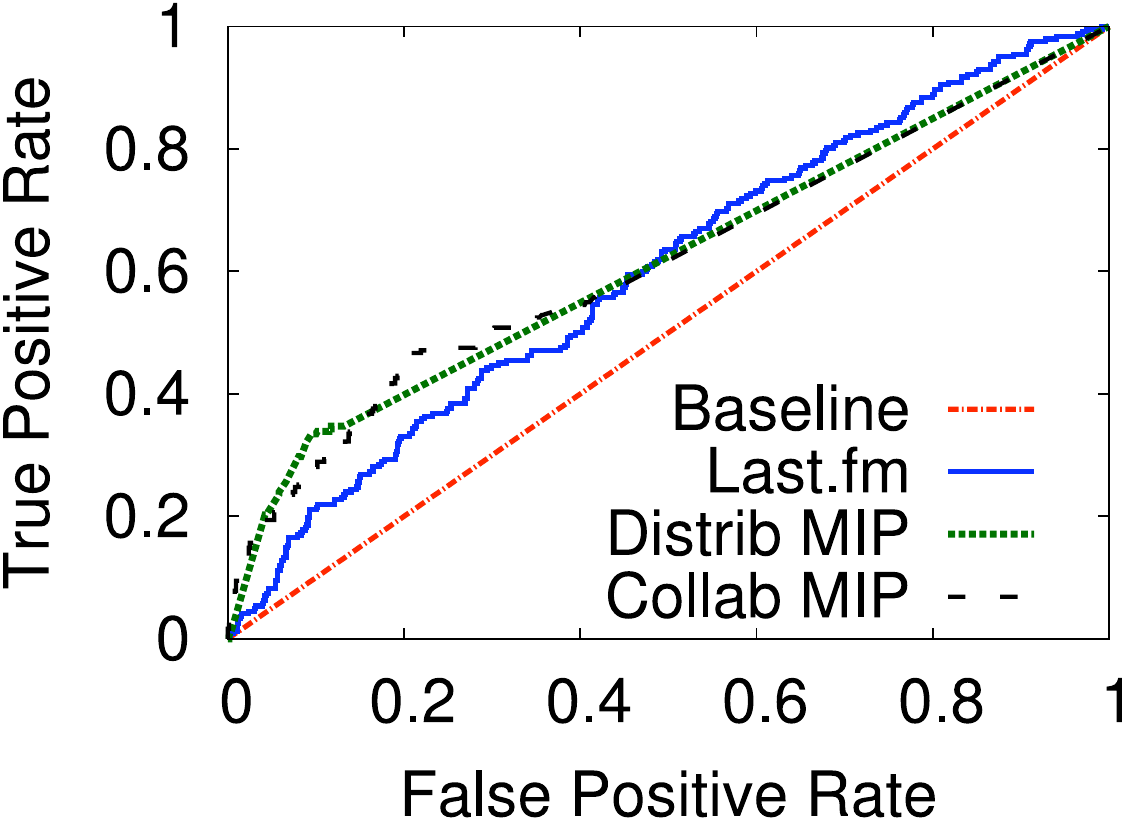} &
\includegraphics[width=0.47\columnwidth]{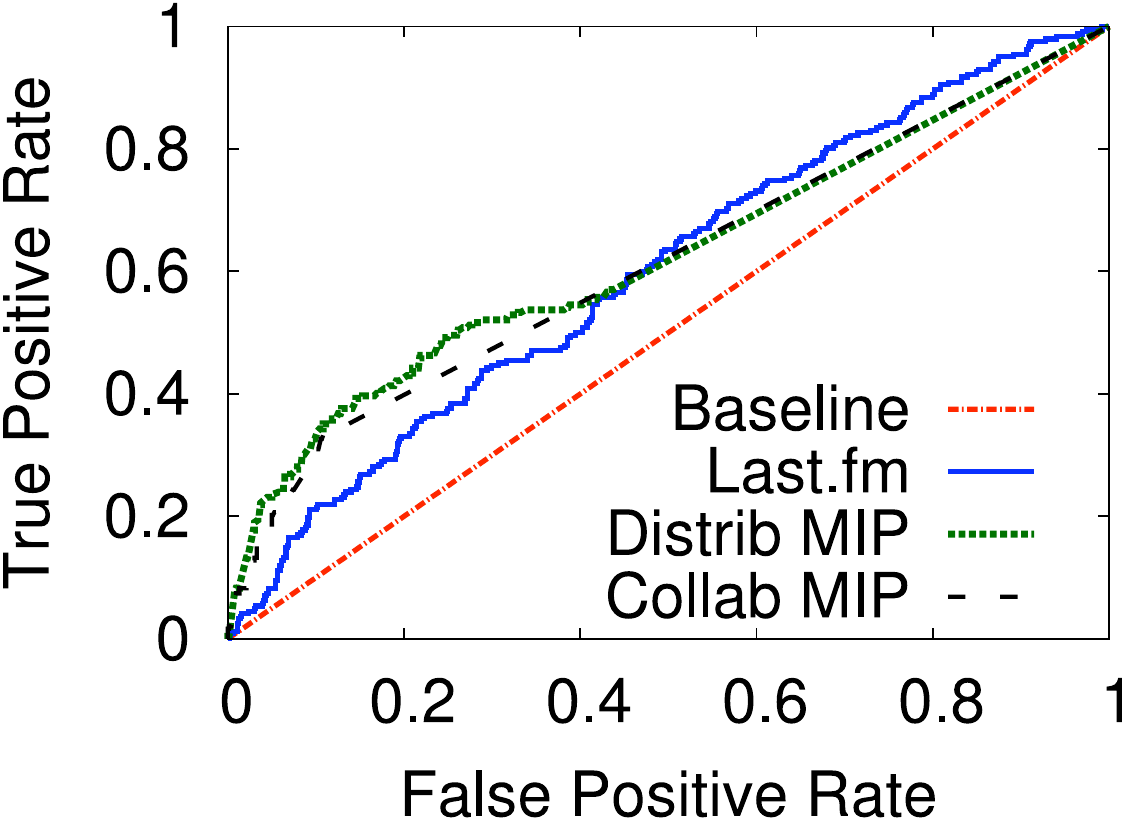} \\
\includegraphics[width=0.47\columnwidth]{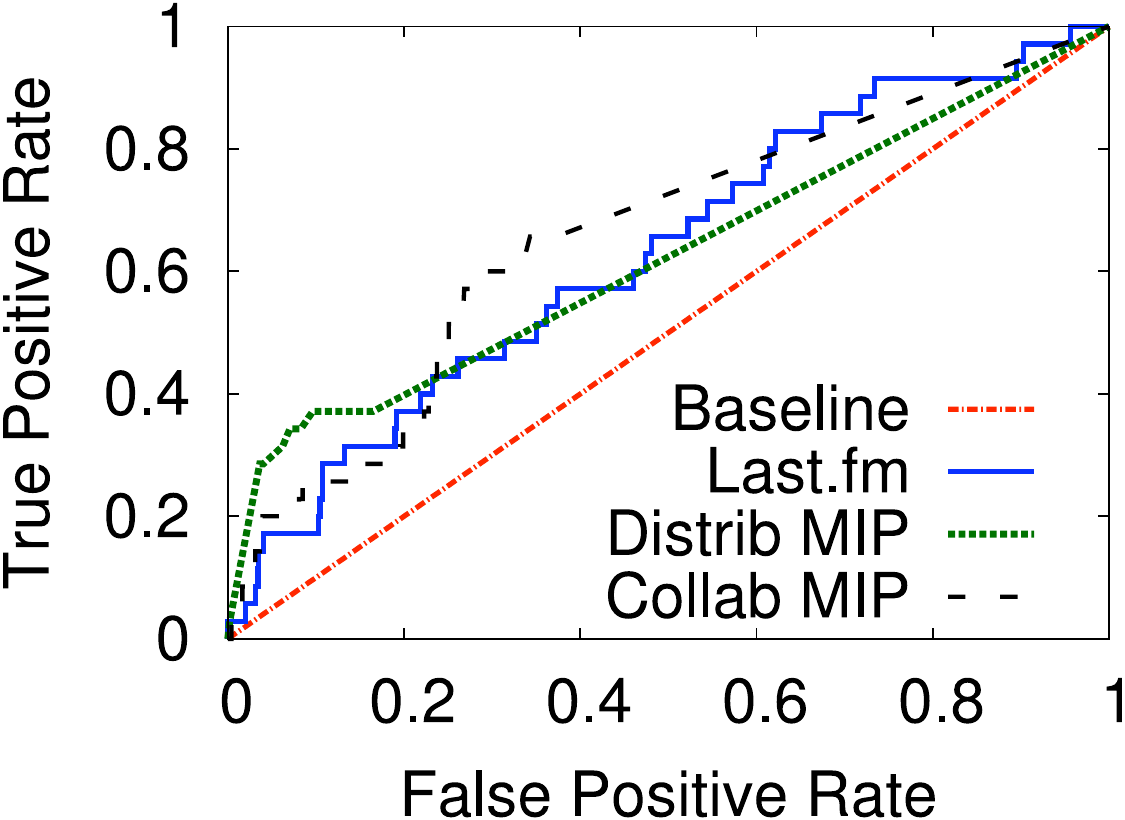} &
\includegraphics[width=0.47\columnwidth]{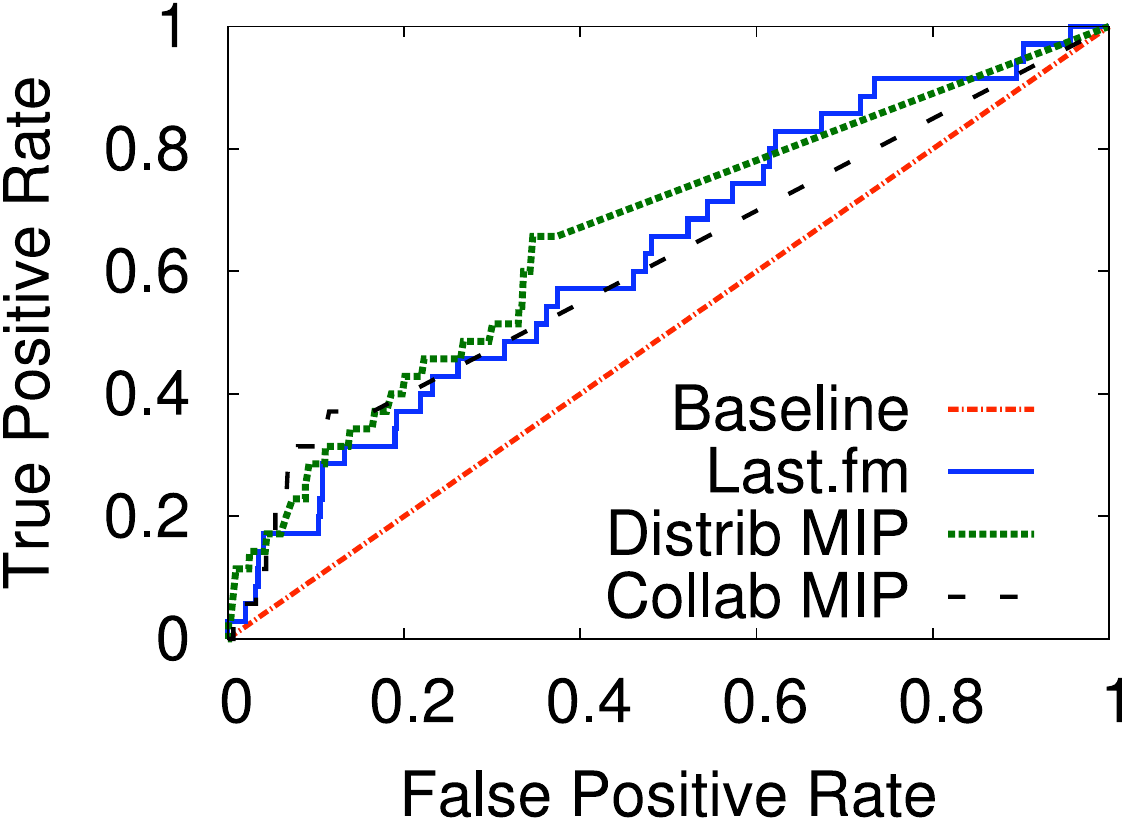}
\end{tabular}
\vspace{-1em}
\caption{ROC curves comparing the social link predictions of distributional and collaborative MIP with the Last.fm recommendations. Triples can be aggregated across items (left) or tags (right). Users are sampled from the most active (top), the most connected (middle), or at random (bottom).}
\label{fig:roc}
\end{figure}

The best results are obtained by sampling the most active users. This is not surprising, as the topical similarity measures have more evidence at their disposal from the users' metadata. In Fig.~\ref{fig:roc} we show ROC plots for the MIP measures, which perform consistently well (among the top 3 measures) in all conditions. While Last.fm neighbor recommendations do perform better than the random baseline, topical similarity is much more accurate than music taste in predicting friends for the most active users. The highest accuracy is achieved by aggregating across items, i.e. representing users as vectors of tags. For the most connected users as well as randomly selected users, the topical similarity measures still perform significantly better than the random baseline, but only marginally better than Last.fm neighbor recommendations. Let us therefore focus on the most active users to evaluate the predictions of additional measures.

\begin{figure}
\centerline{\includegraphics[width=\columnwidth]{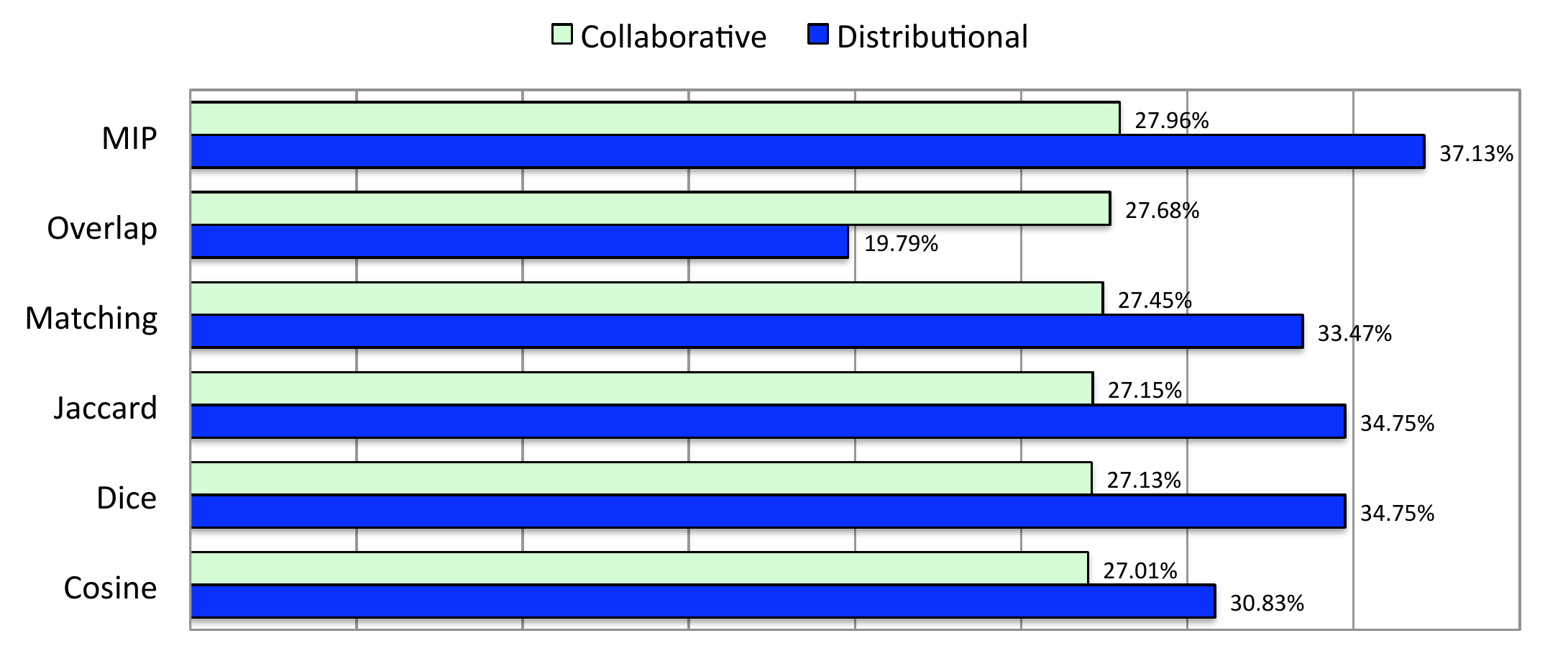}}
\centerline{\includegraphics[width=\columnwidth]{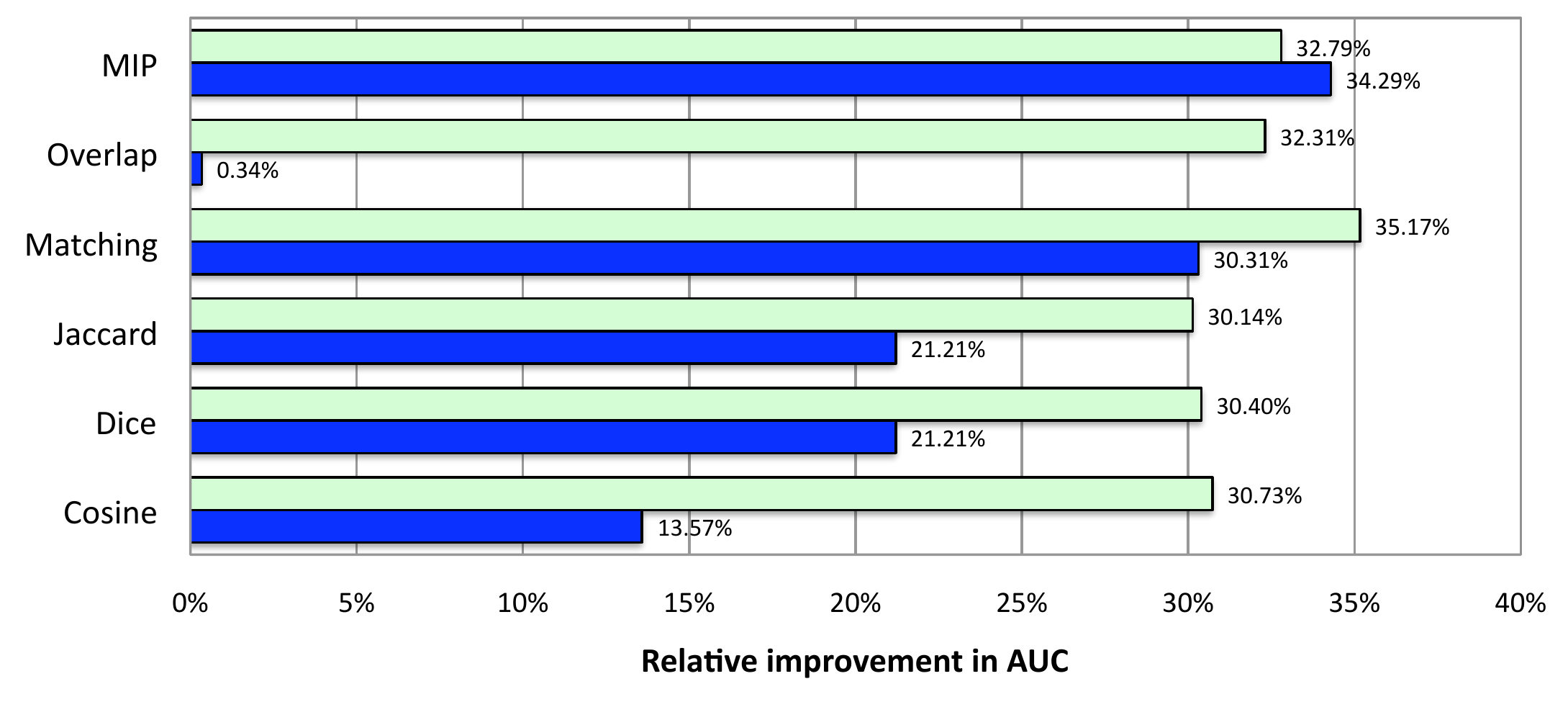}}
\vspace{-1em}
\caption{Relative improvement in area under ROC curves over Last.fm neighbor recommendations, for the most active users. Triples are aggregated across items (top) or tags (bottom).}
\label{fig:auc}
\end{figure}

Since it is difficult to compare 
25 ROC plots, let us summarize our results as follows. For each of the 24 topical similarity measures, $\sigma$, we compare the area under the ROC curve with that obtained by the Last.fm neighbor recommendations. We measure the relative improvement $AUC(\sigma) / AUC(Last.fm) - 1$. A positive number indicates higher accuracy than Last.fm in the sense of a larger number of true positives for the same number of false positives. Fig.~\ref{fig:auc} shows that all topical similarity measures outperform the Last.fm neighbor recommendations. The lonely exception is distributional item overlap, for which the improvement is not significant.  For distributional measures, aggregation across items (focusing on shared tags)  yields better predictions. Overall, the best accuracy is achieved by distributional MIP based on shared tags (37\% improvement). However, if scalability is important, predictions of comparable accuracy can be obtained by projecting over each tag, and then aggregating the similarities across tags. Collaborative matching yields the best predictions in this case (35\% improvement), followed closely by collaborative MIP (33\%) and overlap (32\%).

In summary, these results confirm that the social network constructed from semantic similarity based on user annotations captures actual friendship more accurately than Last.fm's recommendations based on listening patterns. This suggests that the Last.fm neighbourhood selection could be improved by adopting tag-based similarity measures, especially for active users. The results are qualitatively similar for the other sampling methods,  but the differences in accuracy are less significant, with the best predictions outperforming Last.fm by at most 3--4\% in AUC for the most connected users and by 1--5\% for random users. 

%

\section{Conclusion and future work}
\label{sec:conclusions}

In this paper we exploited one peculiarity of Flickr and Last.fm, namely the availability of both
tagging data and the explicit social links between users, to investigate the interplay
of the social and semantic aspects of Web 2.0 applications.

We showed that strong correlations exist between user activity in the social context
(user degree centrality and group participation) and the tagging activity of the same user, and
that a strong assortative mixing exists in the social network; more active nodes
tend to have as neighbors other active nodes.
We also found that a local alignment of users' tag vocabularies is clearly visible 
between nearby users in the social network, even for social tagging systems that 
lack a notion of globally shared tag vocabulary, such as Flickr. 
We investigated the dependence of the number of shared tags and the number of shared
groups of two users, as a function of their shortest-path distance on the social
network. We introduced a null model and we used it to show that part of the similarity
between users who are close on the social network is due to the aforementioned 
correlations between user activity and user degree centrality in the social network. That 
is, assortativity and heterogeneity alone can yield a comparatively higher overlap of
tag usage and group membership for neighboring users.
In this context, our work highlights the importance of backing up the data analysis
with carefully designed null models, which are necessary --- as is the case here --- 
to disentangle the actual signal we are looking for from effects arising purely from 
correlations and mixing properties.

Armed with the null model methodology, we showed that it is possible to define 
measures of tag vocabulary and group membership overlap that are robust with 
respect to the above biases. We investigated the average similarity 
of two users, according to such measures, as a function of the distance in 
the social network, finding that a clear signal of local lexical and topical 
alignment can be detected in Flickr and Last.fm. 

The observed local alignment between lexical (tag) features on the social network 
led us to investigate the question of whether topical similarity measures 
based on social annotations can be applied to the prediction (or recommendation) 
of friend relations in a social network. Last.fm provided us with an ideal 
opportunity to explore this question thanks to the simultaneous availability 
of social link recommendations based on music listening patterns, along with 
the annotation metadata and social network. 

We were able to evaluate the predictive power of a large number of social topical 
similarity measures from the literature, spanning multiple aggregation/projection 
schemes. The results were very encouraging; using any of the tested social 
similarity measures we were able to improve on the accuracy of the social link 
predictions provided by Last.fm, and the improvements were especially significant 
for users who are active taggers. Equally encouraging is the fact that accurate predictions 
are afforded even by incremental measures, pointing to scalable algorithms to 
compute social link recommendations or improve existing methods.

Among the various measures we evaluated, maximum information path has proven 
very accurate across aggregation schemes, data sets, and sampling methodologies. 
When predicting social links between active taggers, MIP is the best measure 
among those based on distributional aggregation (regardless 
of whether we aggregate across items or tags), and either the best or a close 
second among the scalable measures based on collaborative aggregation, 
across items or tags respectively. 

As expected, the Last.fm neighborhood relation seems to be independent of the 
tagging activity of users; we obtain very close AUC values for both the most 
active and most connected sampling strategies. Therefore the number of 
annotations considered does not affect the estimation of user affinity based 
on listening patterns. Accordingly, the present results suggest that the Last.fm 
neighborhood recommendation could benefit considerably from social  
similarity measures --- especially for active users.  

Our results have important implications for the design of social media. 
As social networks and social tagging continue to become increasingly 
popular and integrated in the Web 2.0, our techniques can be directly 
applied to improve the synergies between social and semantic networks --- 
specifically, to help users find friends with similar topical interests 
as well as facilitate the formation of topical communities. 

We plan to further validate our findings via user studies. We will pursue 
this direction by integrating a ``suggest friend'' functionality into 
\url{GiveALink.org}, a social bookmarking system developed by our group 
at Indiana University for research purposes.

On the more theoretical side, future work will consider the present analysis 
performed longitudinally in time, to move from assessing correlations to 
assessing causality. We will investigate whether the activation of a social
link induces a local alignment of tags and group membership, or conversely 
a similarity in interests triggers the creation of a social link. 
Both processes probably play an important role in different situations, 
and adding a temporal dimension to the analysis presented here will provide 
new insight for modeling the structure and evolution of user-driven systems.

%
\subsection*{Acknowledgments}

The authors acknowledge stimulating discussion with A.~Baldassarri, A.~Capocci,
V.~Loreto, and V.~D.~P.~Servedio. We are grateful to Flickr and Last.fm 
for making their data available. 
This work has been partly supported by the \emph{TAGora} project
(\texttt{FP6-IST5-34721}) funded by the FET program of the European Commission 
and by the project 
\emph{Social Integration of Semantic Annotation Networks for Web Applications} 
funded by National Science Foundation award \texttt{IIS-0811994}. 
R.~Schifanella was supported by the World Wide Style project (WWS) of the 
University of Turin.

\end{document}